\documentstyle[11pt,paspconf,epsf]{article}

\newcommand{\kms}{\mbox{\,km\,s$^{-1}$}}
\newcommand{\kpc}{\mbox{\,h$^{-1}$\,kpc}}
\newcommand{\n}{\phantom{0}}

\begin{document}

\title{EFAR peculiar velocities and bulk motions}

\author{Matthew Colless}
\affil{Research School of Astronomy \& Astrophysics, The Australian
       National University, Canberra, ACT 2611, Australia}
\author{R.P. Saglia}
\affil{Institut f\"{u}r Astronomie und Astrophysik, D-81679, Munich, 
       Germany}
\author{David Burstein} 
\affil{Dept of Physics and Astronomy, Arizona State University, Tempe,
       AZ 85287-1504, USA}
\author{Roger Davies}
\affil{Dept of Physics, University of Durham, DH1 3LE, UK}
\author{Robert K McMahan} 
\affil{Dept of Physics and Astronomy, University of North Carolina,
       Chapel Hill, NC 27599-3255, USA}
\author{Gary Wegner}
\affil{Dept of Physics and Astronomy, Dartmouth College, Hanover, 
       NH 03755, USA}

\begin{abstract}
The EFAR project has measured peculiar motions in two distant volumes of
the universe using Fundamental Plane (FP) distances for 85 clusters with
6000\kms$<$$cz$$<$15000\kms. The scatter in distance about the FP is
observed to be 20\% per galaxy for EFAR sample, resulting in peculiar
velocities with a median precision of 1075\kms\ for the 50 clusters with
3 or more galaxies. We find that there is no evidence for a large bulk
flow within either of the two volumes sampled by the EFAR clusters. The
measured bulk motion in both regions is small and consistent with zero.
The Lauer \& Postman (1994) bulk motion is ruled out at the 4$\sigma$
level. There is no evidence supporting the SMAC (Hudson et al.\ 1999) or
LP10K (Willick 1999) bulk motions, but the directionality of the EFAR
sample would only allow a weak (2$\sigma$) detection at best.
\end{abstract}


\section{Introduction}

The main goal of the EFAR project is to use the Fundamental Plane (FP)
for early-type galaxies to measure cluster distances and peculiar
velocities, and so determine the large-scale motions in two relatively
distant volumes. The secondary goals are: (i)~to test for environmental
effects on the cluster FP and evaluate their implications for FP
distance estimates; (ii)~to develop improved distance estimators; and
(iii)~ to study the properties of early-type galaxies in clusters with a
large, homogeneous database.

In this paper we describe the methodology used to fit the Fundamental
Plane and derive peculiar velocities for the clusters in the EFAR
sample. We test the peculiar velocities for systematic errors of various
types, and then examine the velocity field for evidence of bulk motions.
Particular attention is given to testing the claims of large dipoles on
these scales made by Lauer \& Postman (1994), Hudson et al.\ (1999) and
Willick (1998).

\section{EFAR Observations}

The clusters of galaxies in the EFAR sample are selected in two large,
distant (i.e.\ non-local) volumes: Hercules-Corona Borealis (HCB, 40
clusters) and Perseus-Pisces-Cetus (PPC, 45 clusters). The clusters come
from the ACO catalog (Abell et al.\ 1989), the list of Jackson (1982)
and from scans of Sky Survey prints by the authors (Wegner et al.\
1996). The nominal redshift range spanned by the clusters is
6000\kms$<$$cz$$<$15000\kms. The distribution of the EFAR clusters on
the sky, and with respect to the directions of various large-scale
dipoles, is shown in Figure~1.

\begin{figure}
\plotone{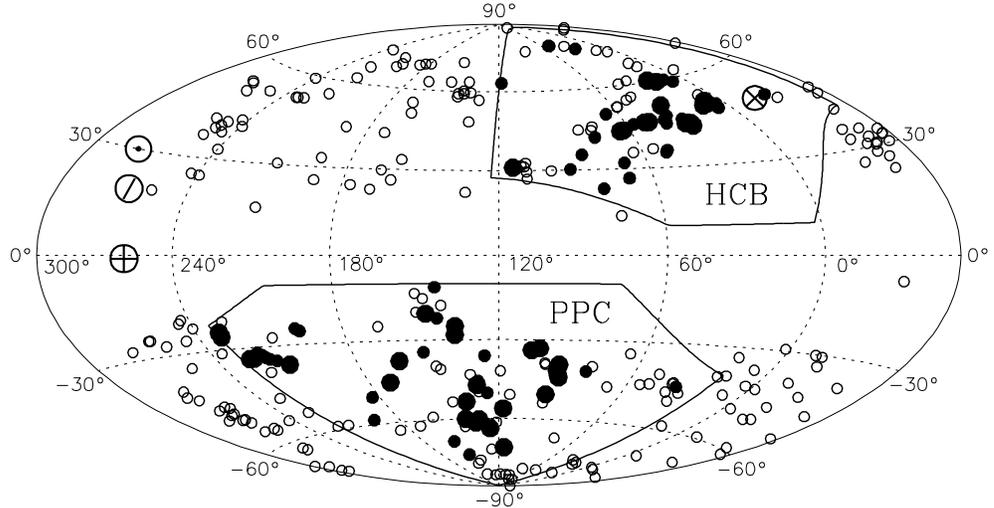}
\caption{The EFAR cluster sample. The plot shows Abell clusters in the
EFAR sample ({\Large $\bullet$}), non-Abell clusters in the EFAR sample
({\small $\bullet$}), and Abell clusters not in the EFAR sample ({\Large
$\circ$}). Also marked are the directions with respect to the CMB frame
of the Local Group dipole ($\odot$), the Lauer \& Postman (1994) dipole
($\otimes$), the SMAC (Hudson et al.\ 1999) dipole ($\oplus$) and the
LP10K (Willick 1998) dipole ($\oslash$).}
\end{figure}

Galaxy selection in each cluster is by elliptical morphology on Sky
Survey prints, and by apparent diameter. The total sample includes 736
early-type galaxies in the 85 clusters. Apparent diameters are measured
visually for all early-type galaxies in the cluster fields. The range in
apparent visual diameter is from 10~arcsec to over 60~arcsec. The sample
selection function is defined in terms of these visual diameters
($D_W$). In total, 2185 $D_W$ diameters were measured for early-type
galaxies in the cluster fields. Selection functions are determined
separately for each cluster, and are approximated by fitted error
functions in $\log D_W$. The mean value of the visual diameter is
$\langle \log D_W \rangle$=1.3 (i.e.\ 20~arcsec), and the dispersion in
$\log D_W$ is 0.3~dex.

Spectroscopic data (redshifts, velocity dispersions and Lick
Mg$b$/Mg$_2$ line indices) were measured for 714 early-type galaxies
from this sample (Wegner et al.\ 1999). A total of 1319 spectra were
obtained on a variety of telescopes, providing repeat measurements for
45\% of the sample. The spectra yielded median errors of 9\% in the
dispersions, 7\% in Mg$b$ and 0.015~mag in Mg$_2$. Comparisons with
literature measurements show good agreement.

Photometry (Saglia et al.\ 1997a) was obtained in the R~band with a
zeropoint uncertainty of 0.03~mag. Circularised galaxy light profiles
were fitted with models having both an $R^{1/4}$ bulge and an
exponential disk. Only 14\% of the sample were adequately fit by a pure
$R^{1/4}$ bulge profile. The uncertainties in the fitted parameters were
estimated from both internal comparisons and simulations (Saglia et al.\
1997b): 90\% of the galaxies have errors in $M_{tot}$ of less than
0.15~mag, in $R_e$ of less than 0.11~dex, and in the FP parameter
$\log R_e-0.3\langle SB_e \rangle$ of less than 0.03~dex.

Membership in the sample clusters (or fore/background clusters) is
determined from redshift distributions using both EFAR redshifts and
redshifts from the ZCAT compilation.

\section{The Fundamental Plane}

The FP is defined from clusters having 6 or more galaxies with reliable
velocity dispersions ($\sigma$), effective radius ($R_e$) and mean
effective surface brightness ($\langle SB_e \rangle$). Simulations show
that including clusters with fewer galaxies increases the variance on
the fitted FP parameters.

To be included in the FP fit, galaxies must have good quality
photometric and spectroscopic data. This means: (i)~high-quality
($Q$=1,2) photometric fits, so $\delta(\log R_e-0.3\langle SB_e
\rangle)$$\le$0.01; (ii)~large and precisely-determined dispersions,
$\sigma$$>$100\kms\ and $\delta\log\sigma$$\le$0.5~dex; (iii)~selection
diameters $D_W$$\ge$12.6\kpc\ and selection probabilities
$P_{sel}$$\ge$0.1. There was {\em no} morphological selection criterion.
After 3$\sigma$-clipping about the FP fit to remove outliers and
interlopers, the final sample comprises 255 galaxies in 29 clusters.

The galaxy distribution in $(\log R_e,\log\sigma,\langle SB_e \rangle)$
space is assumed to be a 3D Gaussian. We perform a simultaneous maximum
likelihood (ML) fit to a global FP and individual peculiar velocity
offsets for each cluster. The parameters of the fit are the FP
coefficients (in the form $\log R_e = a\log\sigma + b\langle SB_e
\rangle + c$), the means and dispersions of the 3D Gaussian galaxy
distribution, and the cluster peculiar velocities ($V_{pec}$). The
peculiar velocities are fitted as offsets $\delta\log R_e$ in the FP,
with $\langle \delta\log R_e \rangle$=0. Residual biases in the peculiar
velocities are removed via simulations. Coma ends up with virtually zero
motion in the CMB frame ($V_{pec}$=$-$29\kms).

The ML Gaussian method has a number of advantages over conventional
approaches to fitting the FP. It accounts for the intrinsic distribution
within the FP (to the extent that this is approximated by a 3D
Gaussian), the errors in all three measured quantities (especially the
relatively large errors in $\sigma$), the {\it a~priori} $D_W$ diameter
selection effects (through selection probability weighting), and the
{\it a~posteriori} selection on $\sigma$, $R_e$ and $P_{sel}$ (by
integrating probabilities over the allowed volume). The method also
minimises the bias in the fitted parameters compared to alternative
regression methods. In detailed simulations comparing the ML Gaussian
method with regression fits, only the ML method gave fitted parameters
with systematic biases smaller than the random errors. It also yielded
the smallest errors in the recovered peculiar velocities, both in terms
of systematic biases and in the scatter in $V_{pec}$ for single
clusters.

\begin{figure}
\plottwo{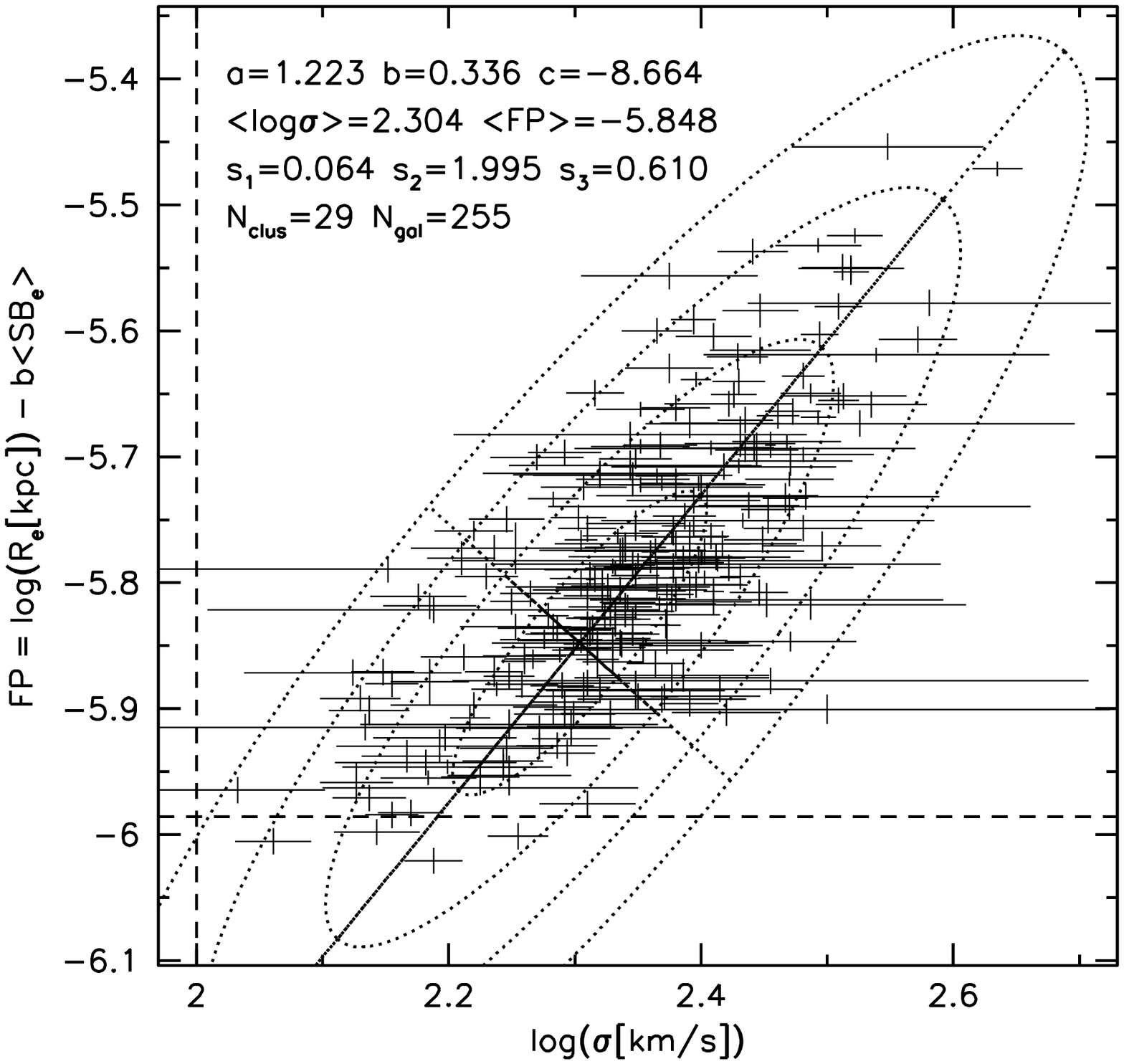}{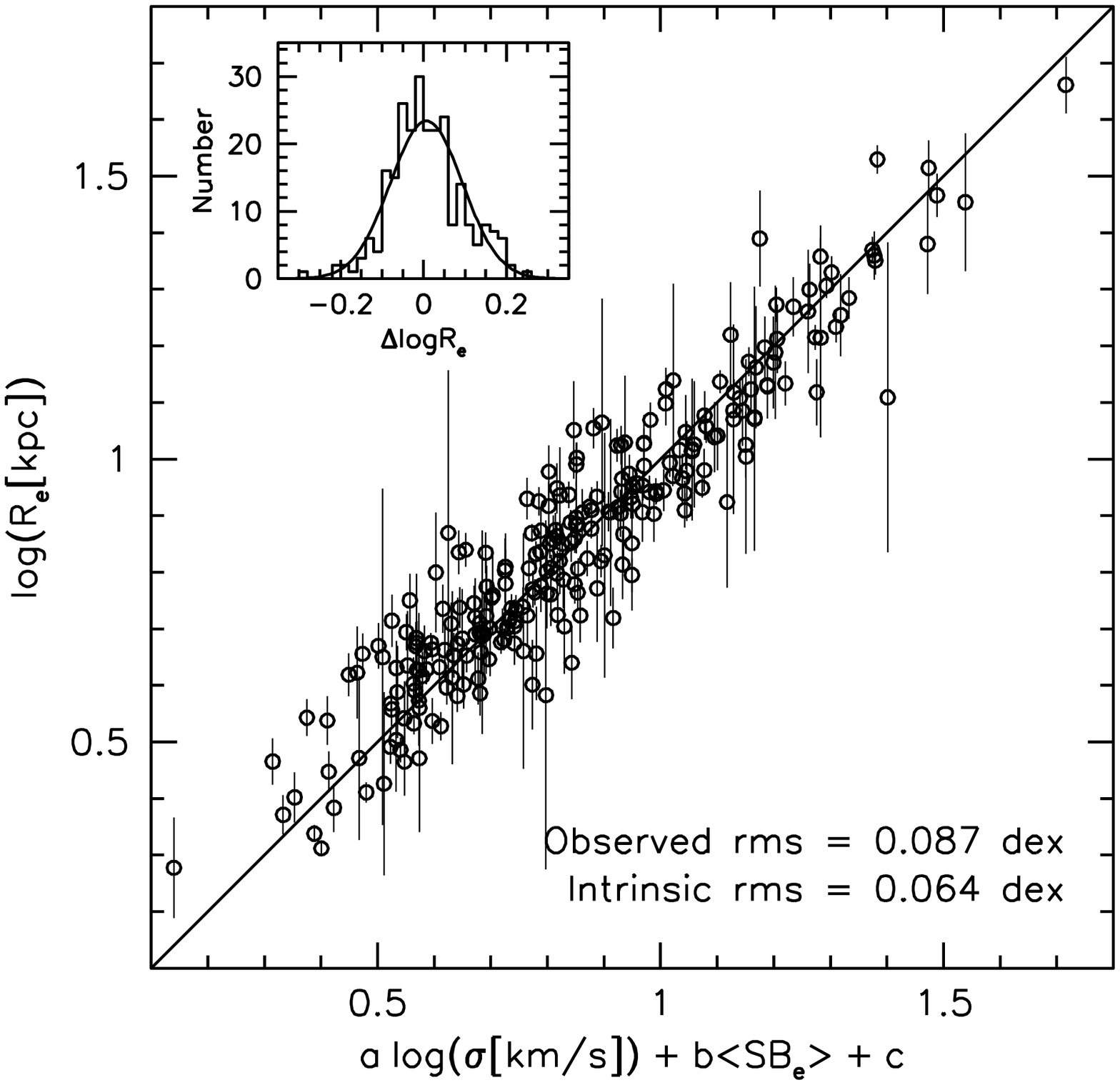}
\caption{The EFAR Fundamental Plane, as fit to the 20 clusters with
$\ge$6 galaxies. The left panel shows the FP projected in the
$(\log\sigma, \log R_e-0.3\langle SB_e \rangle)$ plane and gives the
best-fit parameters; the dotted ellipses are the projected
1$\sigma$-4$\sigma$ contours of the 3D Gaussian fit to the galaxy
distribution. The right panel shows the edge-wise projection of the FP
in the $(\log R_e, a\log\sigma + b\langle SB_e \rangle + c)$ plane and
the distribution of residuals about the fit.}
\end{figure}

Figure~2 shows two views of the best-fit FP for the 20 clusters with
$\ge$6 galaxies. The left panel shows the FP projected in the
$(\log\sigma, \log R_e-0.3\langle SB_e \rangle)$ plane, illustrating the
dominance of the spectroscopic uncertainties over the photometric
uncertainties and giving the parameters of the fit. The right panel
shows the edge-wise projection of the FP in the $(\log R_e, a\log\sigma
+ b\langle SB_e \rangle + c)$ plane and the distribution of residuals
about the fit. The best-fit FP is
\begin{eqnarray}
\log R_e & = & \,\n1.223\log\sigma + 0.336\langle SB_e \rangle - 8.664 . \\
         &   &  \pm0.089\n\n\n\n\,\pm0.013\n\n\n\n\n\pm0.354 \nonumber
\end{eqnarray}
The observed rms scatter about the FP is found to be 0.087~dex (implying
that distance estimates have an uncertainty of 20\% per galaxy), while
the intrinsic scatter (obtained directly from the ML fit of the
intrinsic dispersion about the FP) is 0.064~dex, or 15\%.

The uncertainties in the fitted FP parameters are estimated using 1000
detailed simulations of the fitting process applied to the EFAR dataset.
For all parameters, systematic biases are found to be less than (or at
most comparable to) the random errors. The robustness of the FP fit was
tested by examining 35 variations to the choice of selection and fitting
parameters. Only the most extreme choices (such as applying severe
limits on the galaxies' likelihoods or selection probabilities) gave
significantly different fits. The fit was {\em not} significantly
perturbed by varying the morphological types included, or by the choice
of Burstein \& Heiles (1984) or Schlegel et al.\ (1998) absorption
corrections.

\section{Peculiar Velocities and Bulk Motions}

Once the global FP was determined from the best 20 clusters, peculiar
velocities for {\em all} clusters were obtained by fixing the FP
parameters and applying the ML Gaussian method to determine each
cluster's FP offset, $\delta\log R_e$. In order to minimise the effects
of the uncertainties in the cluster peculiar velocities, we limit the
sample in all subsequent analysis to clusters having at least 3
galaxies, and with $cz$$<$15000\kms\ and $\delta cz_D$$<$2000\kms.

\begin{figure}
\plottwo{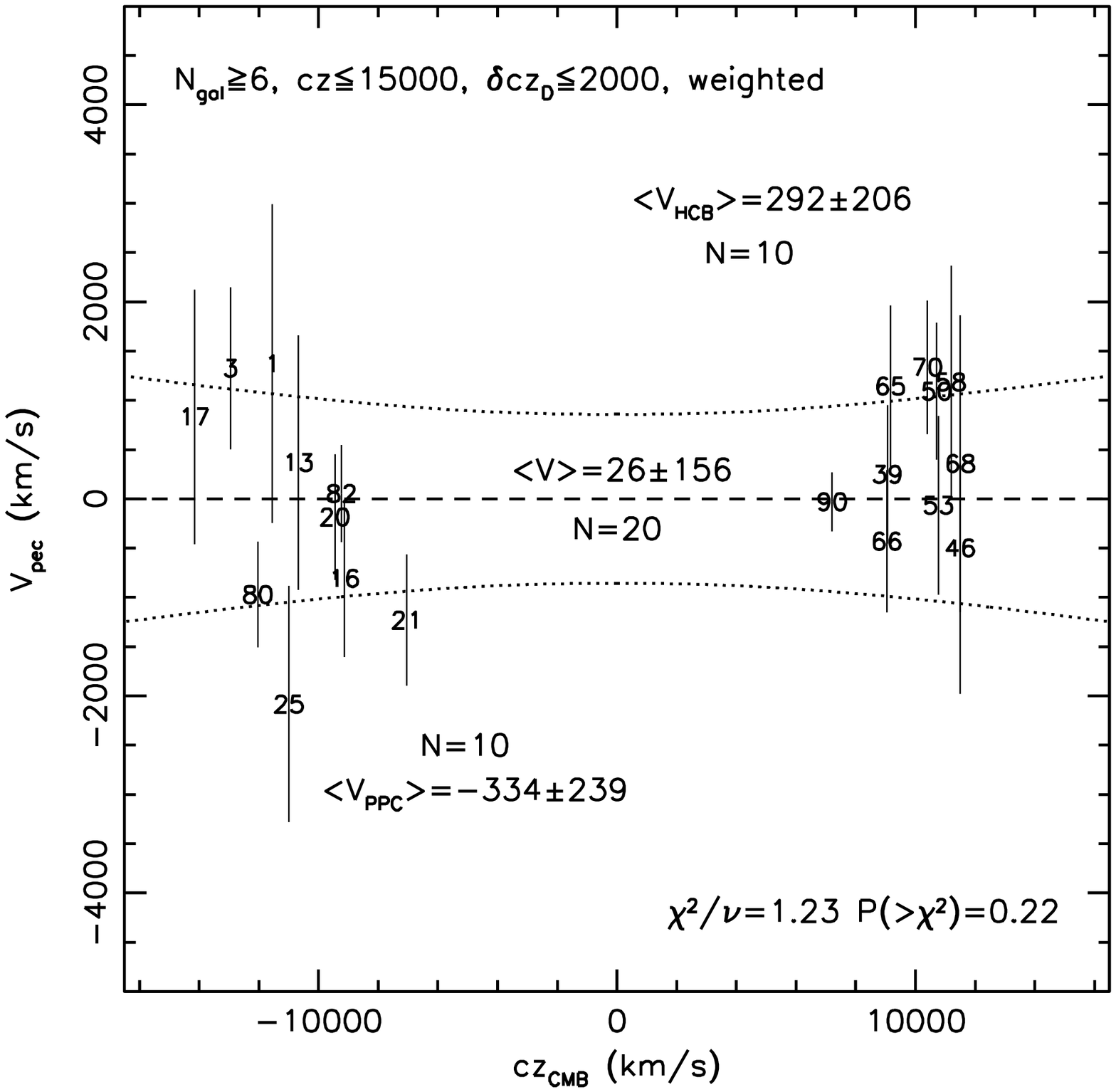}{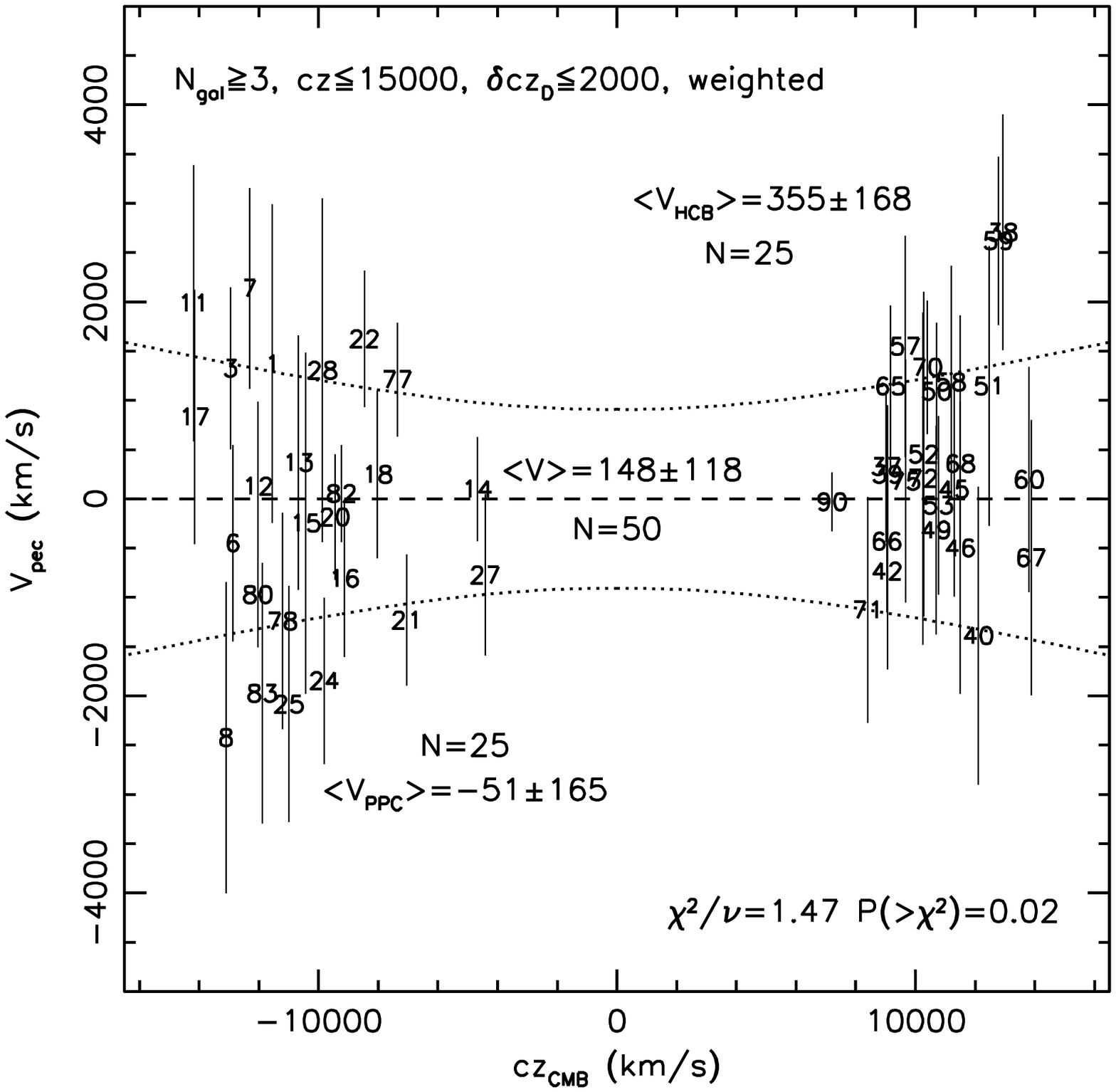}
\caption{Peculiar velocities as a function of redshift. The left panel
shows the 20 clusters with $\ge$6~galaxies; the right panel shows the 50
clusters with $\ge$3~galaxies.}
\end{figure}

We first look to see whether there is any evidence that differences in
the stellar population between clusters are producing systematic errors
in the measured peculiar velocities. We find no correlation between each
cluster's $V_{pec}$ and its offset from the mean cluster Mg--$\sigma$
relation. However, as shown by Colless et al.\ (1999), this is only a
weak test of stellar population effects on FP distance estimates. We
also look to see whether large peculiar velocities could simply be due
to poor FP fits (whether from intrinsic FP variations, observational
errors or cluster interlopers), as has been suggested by Gibbons et al.\
(1999). We find no general trend of increasing $V_{pec}$ amplitude with
the reduced $\chi^2$ of the FP fits, although we do exclude from further
analysis the 3 clusters with $\chi^2/\nu$$>$3, which do have relatively
large $|V_{pec}|$. Finally, we check for radial variations of the mean
peculiar velocity in redshift shells (which might result from a residual
Malmquist bias, for example). We find no evidence for non-zero $\langle
V_{pec} \rangle$ in any direction in any $cz$ shell in either of the two
regions.

\begin{figure}
\plotone{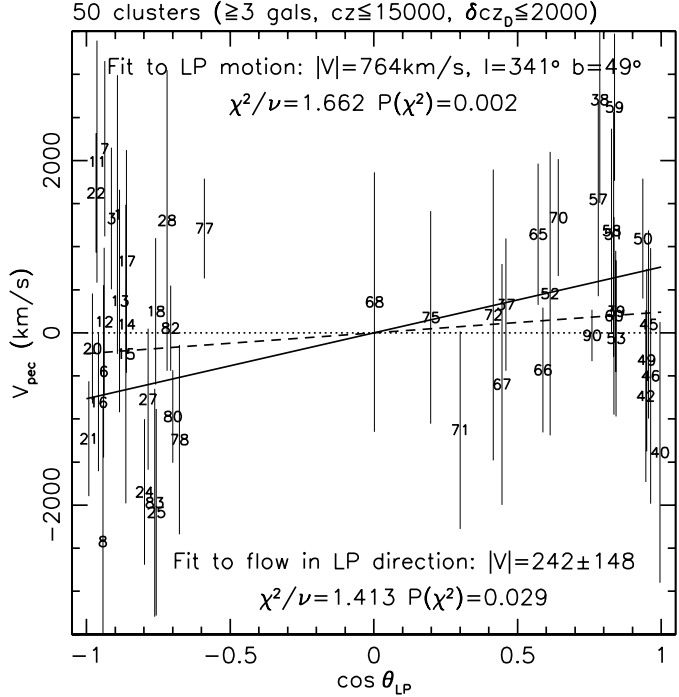}
\caption{Cluster peculiar velocities plotted against the cosine of their
angle with respect to the Lauer \& Postman (1994) dipole (as revised by
Colless 1995). The solid line is the Lauer \& Postman dipole prediction,
which is rejected at the 0.2\% level by a $\chi^2$ test; the dashed line
is a $\chi^2$ fit to a pure bulk motion in this direction, rejected at
the 2.9\% level.}
\end{figure}

With no indication of any systematic biases in our peculiar velocities,
we turn to look for evidence of bulk motions. Figure~3 shows the
peculiar velocities of the clusters as a function of redshift in the CMB
frame, with clusters in the HCB region given positive redshifts and
those in the PPC region given negative redshifts. The left panel shows
the sample of 20 clusters (10 in each of the two regions) with 6 or more
galaxies. There is no evidence for any bulk motion across the sample.
Each region's mean velocity is only non-zero at the 1.5$\sigma$ level,
and a $\chi^2$-test can only reject a model with zero bulk motions at
the 22\% level. If we expand the sample to include all 50 clusters with
3 or more galaxies (again with half in each region), we again find that
the overall motion and the motion in the PPC region are insignificant.
The mean motion of the HCB region, however, is significant at the
2.1$\sigma$ level, and a model with zero bulk motions is rejected by a
$\chi^2$-test at the 2\% level. This apparent motion, however, is
entirely due to just 2 clusters with large $V_{pec}$ (J12, J19), and
disappears if they are omitted from the sample.

As well as searching for a bulk flow in our sample, we can also test
whether the bulk flows obtained by other authors are consistent with our
observed peculiar velocities. Figure~4 shows the peculiar velocities of
the EFAR clusters plotted against the cosine of their angle with respect
to the direction of the bulk flow of 764\kms\ in the direction
$(l,b)$=$(341^\circ,49^\circ)$ found by Lauer \& Postman (1994; as
revised by Colless 1995). This direction is close to the HCB--PPC axis
of the EFAR sample, so our data provide a good test of this claimed bulk
motion. In fact, we find that a $\chi^2$ test rejects the nominal Lauer
\& Postman bulk flow as a representation of the EFAR data at the 0.2\%
level, in accord with the results of Giovanelli et al.\ (1998) and
M\"{u}ller et al.\ (1998). We can perform a similar test to check the
validity of the bulk motion of 630\kms\ in the direction
$(l,b)$=$(260^\circ,-1^\circ)$ claimed by Hudson et al.\ (1999) for the
SMAC cluster sample. However we obtain only a weak constraint because
the SMAC dipole direction is almost orthogonal to the axis of the EFAR
sample, as shown in Figure~1. The same result applies to the LP10K
dipole (Willick 1998), since its direction is close to that of the SMAC
dipole.

\begin{figure}
\plotone{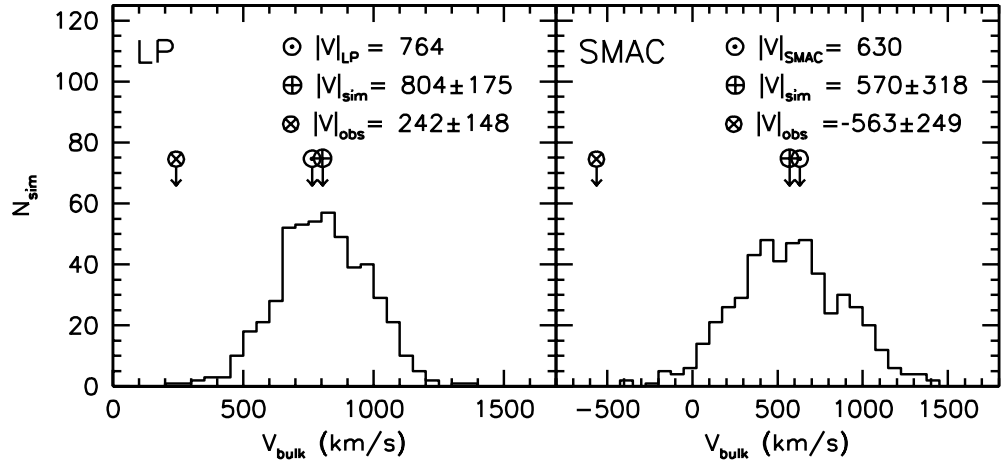}
\caption{Simulation of the bulk motions recovered from the EFAR dataset
assuming the underlying velocity field is described by the Lauer \&
Postman (left) and SMAC (right) bulk motions with no additional
`thermal' peculiar velocities.}
\end{figure}

Figure~5 shows simulations of the bulk motions recovered from the EFAR
dataset under the assumption that the underlying velocity field is
described by the Lauer \& Postman (left) or SMAC (right) bulk flow (with
no additional `thermal' peculiar velocities). The simulations reveal
that EFAR should be able to detect the Lauer \& Postman flow at better
than 4$\sigma$, but that the SMAC flow would not be detectable at better
than 2$\sigma$. These detections would be weaker if there was a
significant thermal component to the peculiar velocity field.

\end{document}